\documentclass[jkps,fleqn,showpacs,twocolumn,showkeys]{revtex4}
\usepackage{graphicx}
\usepackage{amssymb}
\usepackage{amsmath}
\usepackage{bm}
\usepackage{extarrows}
\begin{document}
\title[]{High-precision Estimate of the Critical Exponents for the Directed Ising 
Universality Class}
\author{Su-Chan \surname{Park}}
\email{spark0@catholic.ac.kr}
\thanks{Fax: +82-2-2164-4764}
\affiliation{Department of Physics, The Catholic University of Korea, Bucheon 420-743, Republic of Korea}

\begin{abstract}
With extensive Monte Carlo simulations, we present high-precision estimates of the 
critical exponents 
of branching annihilating random walks with two offspring, a prototypical
model of the directed Ising universality class in one dimension. To estimate the 
exponents accurately, we propose a systematic method to find corrections to scaling
whose leading behavior is supposed to take the form $t^{-\chi}$ in the long-time
limit at the critical point. 
Our study shows that $\chi\approx 0.75$ for the number of particles in defect simulations 
and $\chi \approx 0.5$ for other measured quantities, which should be
compared with the widely used
value of $\chi = 1$. Using $\chi$ so obtained, we analyze the effective exponents
to find that $\beta/\nu_\| = 0.2872(2)$, $z = 1.7415(5)$, $\eta = 0.0000(2)$, and
accordingly, $\beta /\nu_\perp = 0.5000(6)$.
Our numerical results for $\beta/\nu_\|$ and $z$ are clearly different from
the conjectured rational numbers $\beta/\nu_\| = \frac{2}{7} 
\approx 0.2857$, $z = \frac{7}{4}= 1.75$ by Jensen [Phys. Rev. E, {\bf 50}, 
3623 (1994)]. 
Our result for $\beta/\nu_\perp$, however, is consistent with $\frac{1}{2}$,
which is believed to be exact. 
\end{abstract}



\maketitle

\section{INTRODUCTION}
Absorbing phase transitions have been extensively studied during the last several decades. Just 
like equilibrium systems, these non-equilibrium systems are categorized by several 
universality classes according to symmetry and conservation. Notable examples are the 
directed percolation (DP) class and the directed Ising (DI) class, to name only a few (for 
an exhaustive review on universality classes, see, e.g., Refs.~\cite{H2000} and \cite{O2004}). 

Although the deciding features of these universality classes 
are quite well, if not completely, understood (see, for instance, Refs.~\cite{H2000} and \cite{O2004} 
and references therein),
an exact solution of a typical model, even in one dimension, is still not available.
The importance of exact solutions for understanding physical systems in theoretical physics
cannot be exaggerated, and an ample example is the Onsager solution of the two-dimensional 
Ising model~\cite{O1944}. By the same token, any exact solution of a model 
belonging to DP or DI class is still desired.

Since exact solvability is intimately related 
to critical exponents' being rational numbers, whether
the critical exponents obtained from numerical studies can be represented by 
rational numbers has always been a question. For the DP class in one dimension, a certain set of rational numbers 
was proposed as the critical exponents, but
later a detailed numerical analysis clearly disproved that conjecture~\cite{J1996}.
On that account, it does not seem that an exact solution can be found for the DP class.
For the DI universality class, a set of rational numbers for critical exponents
was also conjectured~\cite{J1994}. Although numerical results in the literature look
consistent with the conjectured values within error bars,
whether this conjecture is true or not remains unanswered.

The main aim of this paper is two fold. First, we would like to draw a firm conclusion 
as to whether the critical exponents for the DI class in one dimension are rational
numbers. To state the conclusion first, our extensive numerical study disproves the
conjecture by Jensen~\cite{J1994}. Second, we suggest a numerical method
to extract corrections to scaling that are important for accurate estimates
of the critical exponents. As we will see later, information as to how corrections
to scaling behave at the critical point is crucial when it comes to estimating the critical 
exponents accurately.

This paper is organized as follows: Section~\ref{Sec:model} introduces a
model and defines the quantities in which we are interested.
In Sec.~\ref{Sec:CTS}, we propose a numerical method to find corrections to scaling from
Monte Carlo simulation data. Using this method in Sec.~\ref{Sec:sim}, we find the critical exponents
by studying the corresponding effective exponents.
We summarize our results in Sec.~\ref{Sec:sum}.

\section{\label{Sec:model}Model}
We consider a one-dimensional branching annihilating random walk with two offspring 
(BAW2), which is a prototypical model belonging to the DI class in one dimension. Among many 
versions of the BAW2, we chose the model introduced in 
Ref.~\cite{ZbA1995}. For completeness, we explain below the model and the algorithm we used for 
the simulations.

The BAW2 is defined on a one-dimensional lattice of
size $L$ with periodic boundary conditions. Each lattice
site can be either occupied by a particle ($A$) or be vacant ($\emptyset$).
Double occupancy is not allowed. Each particle can hop to one of its 
nearest neighbors or can branch two offspring.
The detailed dynamic rules are as follows:
\begin{eqnarray}
A\emptyset \xleftrightarrow{(1-p)/2} \emptyset A,\quad
AA \xrightarrow{(1-p)q} \emptyset \emptyset,\nonumber\\
\emptyset A \emptyset \stackrel{p}{\longrightarrow}AAA,\quad
AAA \stackrel{pq}{\longrightarrow}\emptyset A \emptyset ,\quad
AA\emptyset \stackrel{pq}{\longleftrightarrow} \emptyset AA,
\end{eqnarray} 
where the parameters over the arrows represent the corresponding transition rates.

When $q=1$, this model is exactly solvable~\cite{bALR1994} but it does not have 
a nontrivial phase transition in the sense that
for any $p<1$, the density decays to zero as $t^{-1/2}$ with time $t$.
Only when $q$ is smaller than 1, there is a non-trivial critical
point $p_c<1$, and the model belongs to the DI class. In this 
paper, we fix $q=0.5$ (as in Ref.~\cite{ZbA1995}) and investigate the
 behavior of the BAW2 as $p$ varies around the critical point $p_c$. When $p<p_c$ ($p>p_c$),
the system is said to be in the absorbing (active) phase.

In simulations, we used the following algorithm. Assume that there are $N(t)$ particles 
at time $t$ in the system. Among the $N(t)$ particles, a particle 
is chosen at random. It may hop to one of its neighbors (with probability $1-p$)
or may branch two offspring to its two nearest neighbors (with probability $p$).
If a branching attempt is tried with probability $p$, its two neighbors (to be called 
target sites) are examined. 
If both target sites are empty, these two sites become occupied with probability 1.
If one or both of the target sites are already occupied, the branching attempt 
becomes successful only with probability $q$, but with probability $1-q$
this branching attempt is ignored, and nothing happens.
Assume that the branching attempt turns out to be successful.
The empty target site becomes occupied, but the occupied target site becomes empty 
 because of an immediate pair-annihilation event of new and occupied particles. 
In hopping attempts, 
one of two possible directions 
(left or right) is selected at random.
If the selected site is vacant, it lands there with probability $1$. 
If the selected site is occupied, the two particles
undergo pair-annihilation with probability $q$, but the hopping can be ignored
with probability $1-q$, and nothing happens. After an attempt described above,
time increases by $1/N(t)$ regardless of whether the attempt changes the configuration
or not. We repeat the above procedure until
either the system loses all particles or time exceeds 
the preassigned maximum observation time $t_\text{max}$.

In simulations, we used two kinds of initial conditions. 
In one case, simulations begin with the fully-occupied initial condition (FOIC); that is, $N(t=0) = L$. 
In the literature, simulations starting
from the FOIC are generally referred to as static simulations. We will also use
this terminology in this paper.
In the other case, all but a few sites in the middle are empty.
If two consecutive sites are occupied ($N(0) = 2$), this initial condition will 
be called the two-particle initial condition (TPIC).  
We also simulated the stochastic evolution starting from a single particle 
in the whole system ($N(0)=1$), which will be called
the single-particle initial condition (SPIC). Simulations starting from either the TPIC 
or the SPIC will be referred to as defect simulations.
In defect simulations, the system size $L$ should be
large enough so that no particle can hit the boundary up to 
$t_\text{max}$ to ensure that the system size is effectively infinite.

In static simulations, we measured the density $\rho(t) = \langle N(t) \rangle / L$ and
the survival probability $P(t) \equiv \langle 1 - \delta_{N(t),0} \rangle$,
where $\langle \ldots \rangle$ means the average over all realizations and $\delta_{x,y}$
is the Kronecker delta symbol, which should not be confused with the critical exponent 
$\delta$ introduced later. We will also study the density averaged over surviving samples,
$\rho_s(t)$, which is calculated as $\rho_s(t) = \rho(t) / P(t)$.

In defect simulations, we measured the average number of particles
$n(t) \equiv \langle N(t) \rangle$, the survival probability $S(t)\equiv \langle 1 - \delta_{N(t),0} \rangle$, and the square of the distance between the two most distant particles averaged 
over surviving samples, $R^2(t)$. 
Note that we use different symbols for the survival probability
depending on which simulation scheme (static or defect) is under consideration.

\section{\label{Sec:CTS}Scaling relation and corrections to scaling}
According to the scaling theory (for a review, see, e. g., Refs.~\cite{H2000} and \cite{O2004}), 
the asymptotic behavior of $n(t)$, $S(t)$, and $R^2(t)$ 
near the critical point $p_c$ is described as
\begin{eqnarray}
n(t) &\sim& t^\eta f\left (\Delta t^{1/\nu_\|}\right ),\nonumber\\ 
S(t) &\sim& t^{-\delta'} g\left (\Delta t^{1/\nu_\|}\right ), \nonumber\\ 
R^2(t) &\sim& t^{2/z} h\left (\Delta t^{1/\nu_\|}\right ),
\end{eqnarray}
where $\Delta \equiv p - p_c$, $f$, $g$, and $h$ are scaling functions that are not singular
at the origin, and $\eta$, $\nu_\|$, $\delta'$, and $z$ are 
critical exponents. 

When the SPIC is used, $S(t) = 1$ for all $t$ because
of the modulo-2 conservation of the number of particles in the system.
Obviously, $\delta' =0$ in the case of the SPIC.  
In what follows, $S(t)$ exclusively means the survival probability of defect
simulations with the TPIC.  
When the TPIC is used, 
\begin{equation}
\lim_{t\rightarrow \infty} S(t) \sim 
\begin{cases} (p-p_c)^{\beta'}, &p>p_c \text{ (active phase)},\\
0, & p< p_c \text{ (absorbing phase)},
\end{cases}
\end{equation}
which gives the scaling relation $\delta'=\beta'/\nu_\|$.
Likewise, $\rho(t)$ and $\rho_s(t)$ are expected to behave as
\begin{eqnarray}
\rho(t) = t^{-\delta} f_r(t/L^z,\Delta t^{1/\nu_\|}),\nonumber\\
\rho_s(t) = t^{-\delta} g_r(t/L^z,\Delta t^{1/\nu_\|}),
\label{Eq:FSS}
\end{eqnarray}
where $\delta$ is another critical exponent, and $f_r$ and $g_r$ are scaling functions.
Since 
\begin{equation}
\lim_{t\rightarrow \infty} \lim_{L \rightarrow \infty} \rho(t) \sim (p-p_c)^\beta,
\end{equation}
for $p>p_c$, $\delta$ should be equal to $\beta/\nu_\|$.
In general, $\beta$ need not be equal to $\beta'$, but 
the duality relation 
proven in Ref.~\cite{MSS1998} implies
that $\beta$ should be equal to $\beta'$ for the BAW2.

Also, there is the generalized hyperscaling relation~\cite{MDHM1994} 
\begin{equation}
\eta + \delta' + \delta = \frac{d}{z},
\label{Eq:hyper}
\end{equation}
where $d$ is the dimensions in which the system is embedded 
(in this paper, $d$ is always 1).
Note that the above relation is insensitive to the initial conditions of the defect
simulations, although $\eta$ and $\delta'$ individually may be non-universal.

To estimate the critical exponents systematically, one generally uses  the effective exponents 
defined as
\begin{eqnarray}
\eta_\text{eff}(t) &\equiv& \ln \left ( n(t)/n(t/b) \right ) /\ln b,\nonumber\\
-\delta'_\text{eff}(t) &\equiv& \ln \left ( S(t)/S(t/b) \right ) /\ln b,\nonumber\\
\frac{2}{z_\text{eff}(t)} &\equiv& \ln \left ( R^2(t)/R^2(t/b) \right ) /\ln b,\nonumber\\
-\delta_\text{eff}(t) &\equiv& \ln \left ( \rho(t)/\rho(t/b) \right ) /\ln b,
\end{eqnarray}
where $b$ is a constant (in this paper, we set $b=10$). 
In general, there are corrections to scaling, and the measured quantities
at the critical point are expected to behave as
\begin{eqnarray}
n(t) &=& a_n t^{\eta} \left ( 1 + c_n t^{-\chi_n} + o(t^{-\chi_R}) \right ), \nonumber\\
S(t) &=& a_s t^{-\delta'} \left ( 1 + c_s t^{-\chi_s} + o(t^{-\chi_s}) \right ), \nonumber\\
R^2(t) &=& a_r t^{2/z} \left ( 1 + c_r t^{-\chi_r} + o(t^{-\chi_r}) \right ),\nonumber\\ 
\rho(t) &=& a_\rho t^{-\delta} \left ( 1 + c_\rho t^{-\chi_\rho} + o(t^{-\chi_\rho}) \right ), 
\label{Eq:CtoS}
\end{eqnarray}
where $a$'s and $c$'s are constants and $\chi$'s, which will be called the leading
corrections to the scaling exponents (LCSEs), should be positive.
Accordingly, the effective exponents at criticality become
\begin{eqnarray}
\eta_\text{eff}(t) &=& \eta - \frac{c_n}{\ln b} (  b^{\chi_n}-1) t^{-\chi_n} + o(t^{-\chi_n}),\nonumber \\
-\delta'_\text{eff}(t) &=& -\delta' - \frac{c_s}{\ln b} (  b^{\chi_s}-1) t^{-\chi_s} + o(t^{-\chi_s}),\nonumber \\
\frac{2}{z_\text{eff}(t)} &=& \frac{2}{z} - \frac{c_r}{\ln b} (  b^{\chi_r}-1) t^{-\chi_r} + o(t^{-\chi_r}),\nonumber \\
-\delta_\text{eff}(t) &=& -\delta - \frac{c_\rho}{\ln b} (  b^{\chi_\rho}-1) t^{-\chi_\rho} + o(t^{-\chi_\rho}).
\label{Eq:EFF_Def}
\end{eqnarray}
If we plot an effective exponent, for instance $\eta_\text{eff}$, as a function of 
$t^{-\chi_n}$ with the correct value of $\chi_n$, $\eta_\text{eff}(t)$ should
approach the $y$ axis with finite slope at the critical point. Furthermore, if the system
is in the active (absorbing) phase, effective exponents should eventually veer up (down) 
after following a straight line for some time region.
Thus, if the goal of obtaining high-precision estimates of the critical exponents is to be achieved, 
correct information about the LCSE is indispensable.

For models belonging to the DI class,
the LCSE is normally belived to be 1; for an example, see Eq. (5) of Ref.~\cite{J1994}. However, 
numerical data seem to suggest that corrections to scaling for certain quantities 
are stronger than expected. Hence, we feel it necessary to find the 
LCSE more systematically. To this end, we introduce the leading-correction-to-scaling 
function $\Theta_n(t)$ as 
\begin{equation}
\Theta_n(t) \equiv  \left | \frac{n(t) n(t/b^2)}{n(t/b)^2} - 1\right |.
\label{Eq:CTSF}
\end{equation}
In a similar fashion, we can define $\Theta_s(t)$, $\Theta_r(t)$, and $\Theta_\rho(t)$
for $S(t)$, $R^2(t)$, and $\rho(t)$, respectively.
By a straightforward calculation using Eqs.~\eqref{Eq:CtoS} and \eqref{Eq:CTSF}, 
the long time behavior of $\Theta_n(t)$ at criticality becomes
\begin{equation}
\Theta_n(t) = |c_n| \left (b^{\chi_n} - 1\right )^2 t^{-\chi_s}
+ o(t^{-\chi_n}).
\end{equation}
Thus, without prior knowledge of the critical exponents, we can estimate the LCSE
by investigating the leading-correction-to-scaling function.

Although we do not know $p_c$ {\it a priori} in most cases,  
the leading-correction-to-scaling functions and critical
exponents, as well as $p_c$, are obtained at the
same time by using the following iterative procedure:
At first, we make a rough guess about the LCSE, for example, $\chi=1$.
Although the value of the LCSE can be wrong, one can manage to
estimate the critical point with a certain error by observing how a plot of the
effective exponent vs $t^{-1}$ behaves. Now, we calculate the $\Theta$ function and estimate
$\chi$ at the obtained critical point in the above step. Then, with the $\chi$ obtained
from the $\Theta$ function, we re-estimate the 
critical point from longer-time simulations by analyzing plots of
the effective exponents against $t^{-\chi}$. 
At this step, the accuracy of the critical point becomes improved, so
we re-estimate $\chi$ by using the $\Theta$ function at the critical point. 
We repeat the above steps until the resulting values of the $\chi$, $p_c$, and exponents 
become consistent with the scaling theory.
In this paper, we only present the final result of the above procedure.
 
\begin{figure}[t]
\includegraphics[width=0.9\columnwidth]{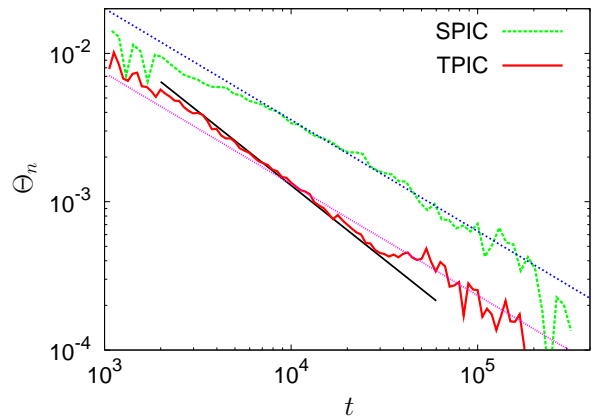}
\caption{\label{Fig:lcse_eta} (Color online) Log-log plots of $\Theta_{n1}(t)$ vs $t$ (upper
curve)  and $\Theta_{n2}(t)$ vs $t$ (lower curve) at
$p = 0.494~675$.
Straight lines with slopes of about $-0.75$ are resulting fitting functions for the 
corresponding curves. A line segment with a slope of $-1$ is also drawn for comparison.}
\end{figure}
\section{\label{Sec:sim}Simulation Results}
This section presents simulations results.  
We begin with analyzing $\Theta_n(t)$ and $\eta_\text{eff}$ from defect simulations
with two different initial conditions.  For convenience, $\eta_\text{eff}$ obtained
from the defect simulations with the SPIC (TPIC) will be denoted by $\eta_1$ ($\eta_2$).
Likewise, $\Theta_n$ with the SPIC (TPIC) will be denoted by
$\Theta_{n1}$ ($\Theta_{n2}$). 

At first, we will show how $\Theta_{n1}$ and $\Theta_{n2}$ behave at $p=0.494~675$,
which we claim to be the critical point of the model.
$\Theta_{n1}$ ($\Theta_{n2}$) is obtained from 
$2  \times 10^8$ ($1.5\times 10^{10}$) independent runs up to $t_\text{max}= 10^{5.5}$ with $b=10$,
but our data for $t > 2\times 10^5$ are too noisy to get reliable information.
Figure~\ref{Fig:lcse_eta} shows double logarithmic plots of 
$\Theta_{n1}(t)$ and $\Theta_{n2}(t)$ against $t$. Both functions are well fitted 
by a power-law function $C t^{-\chi_n}$ with $\chi_n \approx 0.75$ 
in the long-time limit and with $C \approx 3.0$ (1.3) when using the SPIC (TPIC) 
(see the two straight lines in Fig.~\ref{Fig:lcse_eta}). Notice that $\Theta_{n2}(t)$ is 
initially 
well fitted by $t^{-1}$ (see the line segment with a slope of $-1$ in Fig.~\ref{Fig:lcse_eta}),
which might be the reason the effective exponents plotted against $t^{-1}$ in the literature
have given plausible results.

\begin{figure}[t]
\includegraphics[width=0.9\columnwidth]{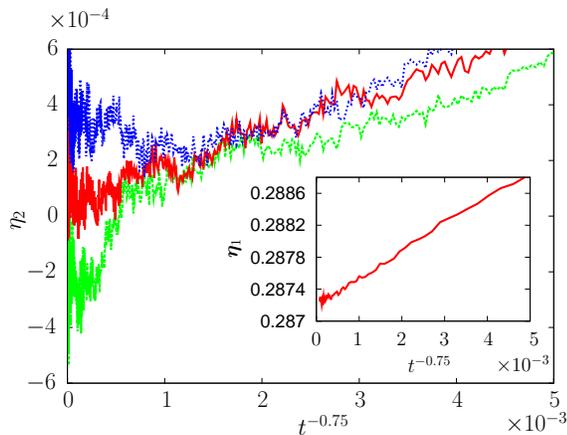}
\caption{\label{Fig:heff2} (Color online) Plots of $\eta_2(t)$ vs $t^{-0.75}$ for $p=0.4947$, $0.494~675$,
and $0.494~65$ (top to bottom). Inset: Plot of $\eta_1(t)$ vs $t^{-0.75}$ 
at $p=0.494~675$.}
\end{figure}

Figure~\ref{Fig:heff2} depicts $\eta_2$ against $t^{-0.75}$ for $p = 0.494~65$, 
$0.494~675$, and $0.4947$ (bottom to top), and
the inset of Fig.~\ref{Fig:heff2} shows the behavior of $\eta_1$ plotted
against $t^{-0.75}$ at $p=0.494~675$. 
For $\eta_1$, we used the same simulation results that were used to calculate
$\Theta_{n1}$ in Fig.~\ref{Fig:lcse_eta},
but for $\eta_2$, we performed other simulations up to $t_\text{max}= 10^7$ with the TPIC.
In this case, the number of independent runs for $p=0.494~65$, $0.494~675$, and $0.4947$ 
were $8\times 10^8$, $8\times 10^8$, and $5\times 10^8$, respectively.
At $p = 0.494~675$, $\eta_2$ becomes a straight line in the region 
where  $t^{-0.75} \le 4\times 10^{-3}$,
but the curve for $p=0.4947$ (0.494~65) veers up (down). Thus, we conclude that
for the TPIC, $p_c = 0.494~675(25)$ and $\eta=0.0000(2)$, where numbers in
parentheses indicate the errors of the last digits.  
Note that our estimate of the critical point is more accurate than that given in 
Ref.~\cite{ZbA1995}. The defect simulation with the SPIC at $p_c$ gives
$\eta_1 = 0.2872(1)$.

\begin{figure}[t]
\includegraphics[width=0.9\columnwidth]{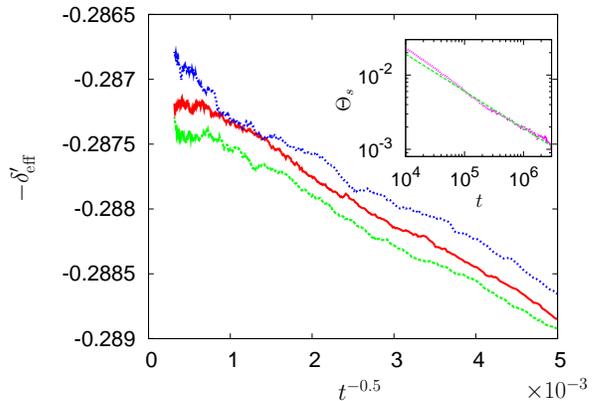}
\caption{\label{Fig:deltaprime} (Color online) Plots of $-\delta_\text{eff}(t)$ vs $t^{-0.5}$ for $p=0.4947$, $0.494~675$,
and $0.494~65$ (top to bottom). Inset: Plot of $\Theta_s(t)$ vs $t$ 
at $p=0.494~675$ on a double-logarithmic scale. The straight line whose slope is
about $-0.5$ is the result of the fitting in the long-time region.}
\end{figure}

Now, we will move to the effective exponent $\delta'_\text{eff}$. In 
Fig.~\ref{Fig:deltaprime}, we depict $-\delta'_\text{eff}(t)$ as a function of
$t^{-0.5}$ for $p=0.4947$, $0.494~675$, and $0.494~65$. 
Because, as shown in the inset of Fig.~\ref{Fig:deltaprime}, $\Theta_s(t)$ behaves
as $\sim t^{-0.5}$ in the long-time limit, 
we plot $-\delta'_\text{eff}$ against $t^{-0.5}$.
By extrapolating $-\delta'_\text{eff}$ for $p = p_c$, we get
$\delta' = 0.2872(2)$. Notice that the corrections to scaling for $S(t)$ are 
stronger than that for $n(t)$.

\begin{figure}[b]
\includegraphics[width=0.9\columnwidth]{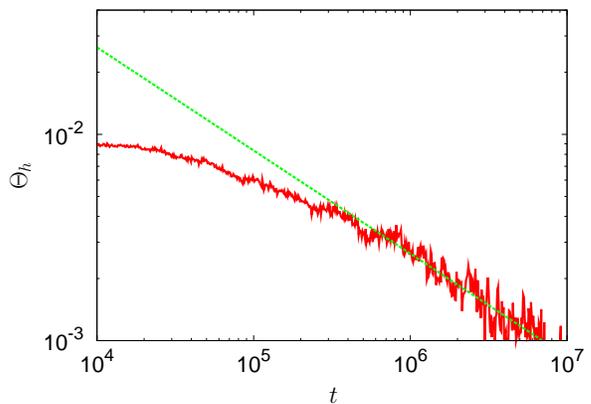}
\caption{\label{Fig:hyper_Q} (Color online) Double-logarithmic plot of $\Theta_h$ 
vs $t$ for $p = 0.494~675$. The straight line with a slope of $-0.5$ is also drawn as a 
guide for the eyes.}
\end{figure}
Finally, we will present the analysis of $z$. It turns out that
the data for $R^2$ are the noisiest among the measured quantities, so it is very hard to 
see a clean asymptotic behavior of $\Theta_r$. Nevertheless we will argue that $\chi_r$, the 
LCSE for $R^2$, is equal to $\chi_s$.  
First note that according to the duality
relation~\cite{MSS1998}, $\rho(t)$ from static simulations and $S(t)$ with the TPIC
should behave in the same way. Thus, the LCSE for $\rho(t)$, that is, $\chi_\rho$, should be
equal to $\chi_s$. Also, according to the hyperscaling relations that are
derived by using the relation
\begin{equation}
\rho(t) \sim \frac{n(t)}{S(t)\sqrt{R^2(t)}},
\label{Eq:s-d}
\end{equation}
both sides of Eq.~\eqref{Eq:s-d} are expected to have the same strengths of corrections to scaling. 
In Fig.~\ref{Fig:hyper_Q}, we depict a double-logarithmic plot of $\Theta_h(t)$ against
$t$, where 
\begin{equation}
\begin{aligned}
\Theta_h &\equiv \frac{H(t)H(t/b^2)}{H(t/b)^2} - 1,\\
H(t)&\equiv \frac{n(t)}{S(t) \sqrt{R^2(t)}}.
\end{aligned}
\end{equation}
Indeed, this function shows a $t^{-0.5}$ behavior in the long-time limit.
Thus, $\chi_r$ should not be smaller than $\chi_s= \chi_\rho$. Hence,
$\chi_r \ge \chi_s$ should be satisfied.
On the other hand, $R^2(t)$ seems to have stronger corrections to scaling than $S(t)$, 
which implies $\chi_r \le \chi_s$. Hence, $\chi_r$ should be equal to $\chi_s$.
\begin{figure}[t]
\includegraphics[width=0.9\columnwidth]{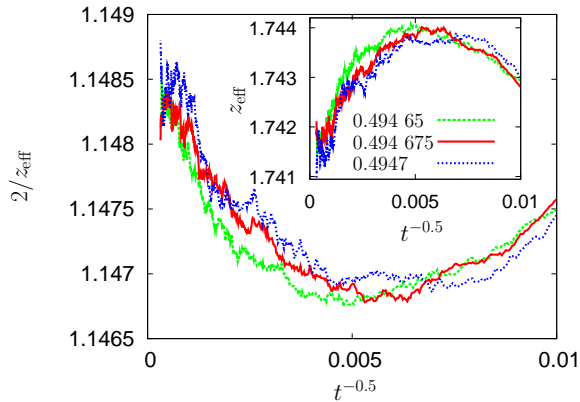}
\caption{\label{Fig:z} (Color online) Plots of $2/z_\text{eff}$ 
vs $t^{-0.5}$ for $p=0.4947$, $0.494~675$, and $0.494~65$. 
Inset: Plots of $z_\text{eff}$ vs 
$t^{-0.5}$ for the same set of $p$'s. $z_\text{eff}$ is defined as
$2/(2/z_\text{eff})$.}
\end{figure}
\begin{figure}[t]
\includegraphics[width=0.9\columnwidth]{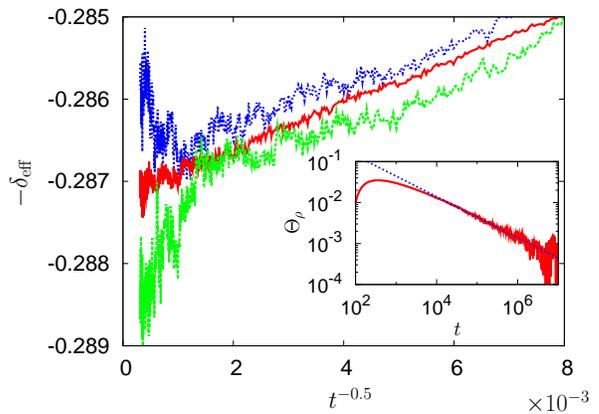}
\caption{\label{Fig:eff} (Color online) Effective exponents $-\delta_\text{eff}$ 
for $p = 0.494~75$, $0.494~675$, and $0.4946$ (top to bottom) as functions 
of $t^{-0.5}$. Inset: Log-log plot of $\Theta_\rho$ against $t$ at $p_c$. The straight line
with a slope of about $-0.5$ is the resulting fitting function of the long-time behavior.}
\end{figure}

In Fig.~\ref{Fig:z}, we plot $2/z_\text{eff}$ against $t^{-0.5}$ near
criticality. From this figure, we conclude that $2/z = 1.1484(4)$ or $z = 1.7415(5)$. See
the inset of Fig.~\ref{Fig:z}, which depicts the behavior of $z_\text{eff} \equiv 2/ (2/z_\text{eff})$.

Next, we will present the results of static simulations. At first, we analyze
$\delta_\text{eff}$ near criticality, as well as the correction-to-scaling function 
$\Theta_\rho$ at the critical point.
In Fig.~\ref{Fig:eff}, we show the behavior of $\delta_\text{eff}$ near
criticality. These curves are obtained from simulations with size $L=2^{23}$ for
the maximum observation time $t_\text{max}=10^7$ at $p = 0.494~75$ (2400 runs), $0.494~675$ 
(10~000 runs), and $0.4946$ (2400 runs) from top to bottom.  
Up to $t_\text{max}$, 
no sample run has fallen into the absorbing state, which minimally supports  the
finite size effect not being significant. Later, we will affirm this statement 
from a finite-size scaling analysis.
The inset of Fig.~\ref{Fig:eff} gives the reason $-\delta_\text{eff}$ is
plotted against $t^{-0.5}$; in the long-time limit, $\Theta_\rho$ decays as $t^{-0.5}$ at $p=p_c$.
Hence, we conclude that $\delta = 0.2872(1)$, which
is consistent with the duality relation $\delta = \delta'$.
Also, note that within the error bars, our exponents are consistent with the hyperscaling relation 
Eq.~\eqref{Eq:hyper}.
In particular, if $\eta_2$ is exactly 0 (see Fig.~\ref{Fig:heff2}), we see that
$2 \delta = 1/z$ or $\delta z \equiv \beta/\nu_\perp = \frac{1}{2}$, exactly. 

\begin{figure}[b]
\includegraphics[width=0.9\columnwidth]{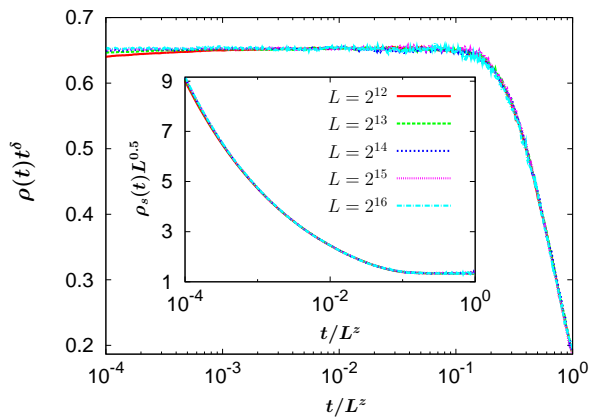}
\caption{\label{Fig:Sfss} (Color online) Scaling collapse plots of $\rho(t) t^\delta$ vs $t/L^z$
with $\delta = 0.2872$ and $z=1.7415$ for $L=2^{12}$, $2^{13}$, $2^{14}$, $2^{15}$,
and $2^{16}$ at $p=p_c$ on a semi-logarithmic scale. Inset: Scaling collapse plots of 
$\rho_s(t) L^{0.5}$ vs $t/L^z$ for the same system sizes at criticality on a semi-logarithmic scale.}
\end{figure}
Figure~\ref{Fig:Sfss} depicts the finite-size scaling collapse at the critical
point. According to the scaling ansatz in Eq.~\eqref{Eq:FSS}, plots of $\rho(t) t^\delta$ vs
$t /L^z$ for different $L$'s at the critical point should collapse into a single curve.
Indeed, this scaling collapse is clearly observed in Fig.~\ref{Fig:Sfss} when 
we use the critical exponents obtained above. Also, the inset of Fig.~\ref{Fig:Sfss} 
clearly shows the scaling collapse of plots of $\rho_s(t) L^{0.5}$ vs $t/L^z$,
which is consistent with $\beta/\nu_\perp = \frac{1}{2}$, obtained from the hyperscaling
relation.  Because the finite-size effect becomes significant when $t \ge 0.1 \times L^z$, as can be 
deduced from Fig.~\ref{Fig:Sfss}, we expect
the finite-size effect for $L = 2^{23}$ to become crucial when $t \ge 10^{12}$,
which is much larger than the maximum observation time $10^7$ in Fig.~\ref{Fig:eff}. 
Thus, we did not have to take the finite-size effect into account 
when we analyzed $\delta_\text{eff}$ in Fig.~\ref{Fig:eff}.

\section{\label{Sec:sum}Summary}
To sum up, we presented high-precision estimates of the critical exponents 
for the branching annihilating random
walks with two offspring, which is a prototypical model belonging to the directed
Ising universality class in one dimension. 
To this end, we first analyzed corrections to scaling by using the correction-to-scaling
functions defined in Eq.~\eqref{Eq:CTSF}. This method can be easily applicable
to any critical systems, although reducing statistical fluctuations by
simulating many independent runs is the main practical obstacle.
From this analysis, we found that the LCSE for $n(t)$ 
was about $0.75$, but those for other measured quantities were all around $0.5$.
With the LCSE obtained, we analyzed the effective exponents and found that
$\eta = 0.0000(2)$ and $\delta' = 0.2872(2)$ when the two-particle initial condition was used
and that $\eta = 0.2872(1)$ and $\delta' = 0$ when the single-particle initial condition was used.
We also found that $z= 1.7415(5)$ and $\delta = 0.2872(1)$. These exponents
distinctively differ from the conjectured rational numbers $z =\frac{7}{4} = 1.75$
and $\delta = \frac{2}{7} \approx 0.2857$ in Ref.~\cite{J1994}. If $\eta$ is
exactly zero, the generalized hyperscaling relation, along with the duality property
of the BAW2, shows that $\beta/\nu_\perp = \frac{1}{2}$ exactly, and our numerical simulations
are consistent with this value within the error bars. All the numerical results
are summarized in Table~\ref{Table:exp}, along with the conjectured values for comparison.
\begin{table}
\caption{\label{Table:exp}Critical exponents of the BAW2 in one dimension.}
\begin{ruledtabular}
\begin{tabular}{ccc}
exponents & simulations & conjecture~\cite{J1994} \\
\colrule
 $\eta$ & 0.0000(2) & 0 \\
 $\beta/\nu_\|$ & 0.2872(2) & $\frac{2}{7}\approx 0.2857$ \\
 $z$ & 1.7415(5) & $\frac{7}{4}$ \\
 $\beta/\nu_\perp$ & 0.5000(6) & $\frac{1}{2}$ \\
\end{tabular}
\end{ruledtabular}
\end{table}

\begin{acknowledgments}
This work was supported by the Catholic University of Korea, Research Fund 2011 and
by the Basic Science Research Program
through the National Research Foundation of Korea
(NRF) funded by the Ministry of Education, Science
and Technology (Grant No. 2011-0014680).
\end{acknowledgments}


\begin{thebibliography}{99}
\bibitem{H2000} H. Hinrichsen, Adv. Phys. {\bf 49}, 815 (2000).
\bibitem{O2004} G. {\'O}dor, Rev. Mod. Phys. {\bf 76}, 663 (2004).
\bibitem{O1944} L. Onsager, Phys. Rev. {\bf 65}, 117 (1944).
\bibitem{J1996} I. Jensen, J. Phys. A {\bf 29}, 7013 (1996).
\bibitem{J1994} I. Jensen, Phys. Rev. E {\bf 50}, 3623 (1994). 
\bibitem{ZbA1995} D. Zhong and D. ben-Avraham, Phys. Lett. A {\bf 209}, 333 (1995). 
\bibitem{bALR1994} D. ben-Avraham, F. Leyvraz, and S. Redner, Phys. Rev. E {\bf 50}, 1843 (1994). 
\bibitem{MSS1998} K. Mussawisade, J. E. Santos, and G. M. Sch{\"u}tz, J. Phys. A {\bf 31}, 4381 (1998).
\bibitem{MDHM1994} J. F. F. Mendes, R. Dickman, M. Henkel, and M. C. Marques, J. Phys. A {\bf 27}, 3019 (1994).
\end{thebibliography}
\end{document}